\documentclass[pdflatex,sn-nature]{sn-jnl}

\usepackage{amsmath,amssymb,amsfonts} 
\usepackage{graphicx}                 
\usepackage{relsize}                  
\usepackage{xcolor}                   
\usepackage{units}                    
\usepackage{epsfig}                   
\usepackage[title]{appendix}          
\usepackage{bm}                       
\usepackage{float}                   
\usepackage{lineno}

\begin{document}

\title[Article Title]{Resolving electronic evolution during bond dissociation}

\author*[1,2]{\fnm{Tian} \sur{Wang}}\email{twang110@uottawa.ca}
\equalcont{These authors contributed equally to this work.}
\author[1]{\fnm{Nida} \sur{Haram}}
\equalcont{These authors contributed equally to this work.}
\author[1]{\fnm{Zack} \sur{Dube}}
\author[3]{\fnm{Kyle A.} \sur{Hamer}}
\author[1]{\fnm{Yonghao} \sur{Mi}}
\author[1]{\fnm{Fatemeh} \sur{Karimi}}
\author[1]{\fnm{Andrei Yu.} \sur{Naumov}}
\author[1]{\fnm{Giulio} \sur{Vampa}}
\author[4]{\fnm{Caterina} \sur{Vozzi}}
\author[2]{\fnm{Xiaojun} \sur{Liu}}
\author[1]{\fnm{Albert} \sur{Stolow}}
\author[5]{\fnm{Michael S.} \sur{Schuurman}}
\author[3]{\fnm{Nicolas} \sur{Douguet}}
\author[1]{\fnm{David M.}  \sur{Villeneuve}}
\author[1]{\fnm{Paul B.} \sur{Corkum}}
\author*[1]{\fnm{Andr\'{e}} \sur{Staudte}}\email{andre.staudte@nrc-cnrc.gc.ca}

\affil[1]{\orgdiv{Joint Center for Extreme Photonics}, \orgname{National Research Council and University of Ottawa}, \orgaddress{\city{Ottawa}, \state{Ontario}, \country{Canada}}}
\affil[2]{\orgdiv{State Key Laboratory of Magnetic Resonance and Atomic and Molecular Physics}, \orgname{Wuhan Institute of Physics and Mathematics, Innovation Academy for Precision Measurement Science and Technology, Chinese Academy of Sciences}, \orgaddress{\city{Wuhan}, \postcode{430071}, \country{China}}}
\affil[3]{\orgdiv{Department of Physics}, \orgname{University of Central Florida}, \orgaddress{\city{Orlando}, \state{FL}, \country{USA}}}
\affil[4]{\orgdiv{CNR IFN - Istituto di Fotonica e Nanotecnologie}, \orgaddress{\street{Piazza Leonardo da Vinci 32}, \city{Milano}, \postcode{20133}, \country{Italy}}}
\affil[5]{\orgdiv{National Research Council Canada}, \orgaddress{\city{Ottawa}, \state{Ontario}, \postcode{K1A 0R6}, \country{Canada}}}


\abstract{Coupled electronic and nuclear motions govern chemical reactions, yet resolving how electronic structure evolves during bond dissociation remains a central challenge. Here we investigate the photodissociation of Br$_2$ using correlated photoelectron–photoion coincidence measurements. A 400~nm pulse initiates dissociation, while strong-field ionization probes the evolving molecular system. Coincident measurement of three-dimensional photoion and photoelectron momenta provides simultaneous access to the internuclear separation and the accompanying electronic evolution. We identify multiple distinct stages of electronic evolution during bond dissociation. The reshaping of the ionizing molecular orbital occurs first, followed by redistribution and localization of the electronic charge density, and finally by the gradual decay of residual electronic coherence between the separating atomic centers. Between the molecular and atomic limits, we observe an intermediate bond-breaking state in which localized atomic character coexists with a partially delocalized electronic response. By combining correlated observables with semiclassical modelling, we resolve the temporal ordering of these coupled electronic and nuclear processes and determine their associated dynamical timescales. These results demonstrate how correlated momentum observables can disentangle different aspects of the molecular-to-atomic transition.
}

\maketitle

\section*{Main}
Understanding how a chemical bond evolves and ultimately breaks during a reaction is a central task in chemistry \cite{Zewail1988,Gessner2006Science,Li2010PNAS,Leone2019}. Experimentally, this remains challenging because the temporal ordering of distinct changes in electronic structure relative to nuclear motion is not directly accessible from any single observable. This ambiguity underlies the absence of a precise criterion for identifying when a chemical bond ceases to exist \cite{Hait2023}. It also reflects the broader challenge of defining dissociation times in photochemical reactions such as Br$_2$, where different observables probe different stages of the evolving electron-nuclear dynamics \cite{Beetar2025JPCA}. Here, we focus on this prototypical example of bond dissociation in Br$_2$, where absorption of a 400~nm photon excites the molecule to a dissociative electronic state, ultimately yielding two neutral ground-state atoms \cite{Cooper1998}.

In a strict quantum-mechanical description, bond breaking is not associated with a sharply defined internuclear distance or a unique moment in time. Instead, it corresponds to a gradual transition from a regime in which electronic structure is delocalized over multiple atomic centers to one in which the electronic structure becomes effectively atomic and interatomic coupling is strongly reduced. Hence, the “dissociation time” is not a single instant, but a characteristic timescale over which the nuclear wave packet evolves toward separated atomic fragments. A key aspect of this evolution is the loss of electronic coherence, defined here as the gradual reduction of coherent coupling between electronic amplitudes on different atomic centers. Thus, the system necessarily passes through an intermediate regime in which this coherence is only partially lost. The electronic structure then exhibits both molecular character, due to the remaining interatomic coupling, and atomic character, reflecting the emerging electronic localization on individual atoms as this coupling diminishes. We refer to this transitional regime as the “bond-breaking state” (BBS).

Experimentally resolving these coupled dynamics requires simultaneous access to both the evolving nuclear configuration and the electronic structure. Previous single-observable approaches, including photoelectron spectroscopy \cite{Nugent2001PRL, Wernet2009PRL}, high-harmonic generation spectroscopy \cite{Worner2010}, and strong-field ion momentum spectroscopy \cite{Li2010PNAS, Rouzee2013JPB}, have provided valuable insights into either electronic or nuclear dynamics, or have inferred their interplay indirectly. Electron-ion coincidence techniques have enabled correlated measurements of electronic and nuclear degrees of freedom in molecular systems (e.g., \cite{Landers2001PRL, Weber2004Nature, Jahnke2010NatPhys, Hockett2011NatPhys, Li2021SciRep}), including kinematically complete measurements of molecular fragmentation dynamics. However, directly resolving how distinct electronic observables evolve relative to nuclear separation during bond dissociation remains challenging. In particular, establishing the temporal ordering of electronic restructuring, charge localization, and fragment formation during dissociation has remained largely unexplored.

Here, we perform coincident measurements of photoelectron and photoion momenta following strong-field ionization of dissociating Br$_2$ using COLd Target Recoil Ion Momentum Spectroscopy (COLTRIMS) \cite{Ullrich2003}. We monitored the dissociation of Br$_2$ using a two-colour pump–probe scheme in which a 400~nm pulse excites the molecule from the X $^1\Sigma_g^+$ ground state to the dissociative C $^1\Pi_u$ state (Br$_2^*$), followed by ionization with a time-delayed 800~nm pulse (see Fig.~\ref{fig1}a). Owing to the perpendicular ($\Sigma\rightarrow\Pi$) character of the transition, molecules with their internuclear axis oriented perpendicular to the laser polarization are preferentially excited. The subsequent neutral dissociation therefore occurs predominantly in the plane perpendicular to the polarization direction. Both pulses were linearly polarized along the same axis, ensuring a well-defined probing geometry (as illustrated in Fig.~\ref{fig1}).

Interaction of the delayed 800~nm pulse with Br$_2^*$ produces both single- and double-ionization outcomes. In the single-ionization channel, one atom is ionized while the other remains neutral, whereas in the double-ionization channel both atoms are ionized. By measuring the three-dimensional momenta of all photoions and photoelectrons in coincidence, these channels can be unambiguously separated, and the corresponding electrons assigned to their respective molecular pathways. This correlated coincidence approach provides simultaneous access to the evolving nuclear separation and the changing electronic structure, allowing us to directly follow coupled electron-nuclear dynamics throughout the dissociation.

Our results reveal that the electronic evolution proceeds through multiple distinct stages, including reshaping of the orbital, redistribution and localization of electronic charge, and the gradual loss of coherence between the two atomic centers. Between the molecular and atomic limits, an intermediate regime emerges in which localized atomic character coexists with a partially delocalized electronic response, corresponding to the bond-breaking state introduced above. These electronically distinct stages occur at different regions along the dissociation coordinate and manifest as characteristic structures within the photoelectron momentum distributions, enabling multiple aspects of the evolving electronic structure to be resolved simultaneously.

Importantly, this multi-stage evolution goes beyond a simple separation of electronic and nuclear dynamical timescales. Instead, it reflects a structured sequence of electronic processes that unfold during bond dissociation and are governed by electron-nuclear coupling. By combining correlated observables with semiclassical modelling, we provide an experimental approach to resolve the temporal ordering of electronic dynamics during molecular dissociation. More broadly, these results demonstrate how correlated photoelectron and photoion measurements can simultaneously resolve multiple electronic observables during molecular dissociation, offering new insight into how chemical bonds evolve and ultimately dissociate.

\section*{Results}

\subsubsection*{Correlated photoion-photoelectron spectra}

\begin{figure}[H]
    \centering
    \includegraphics[width=0.9\textwidth]{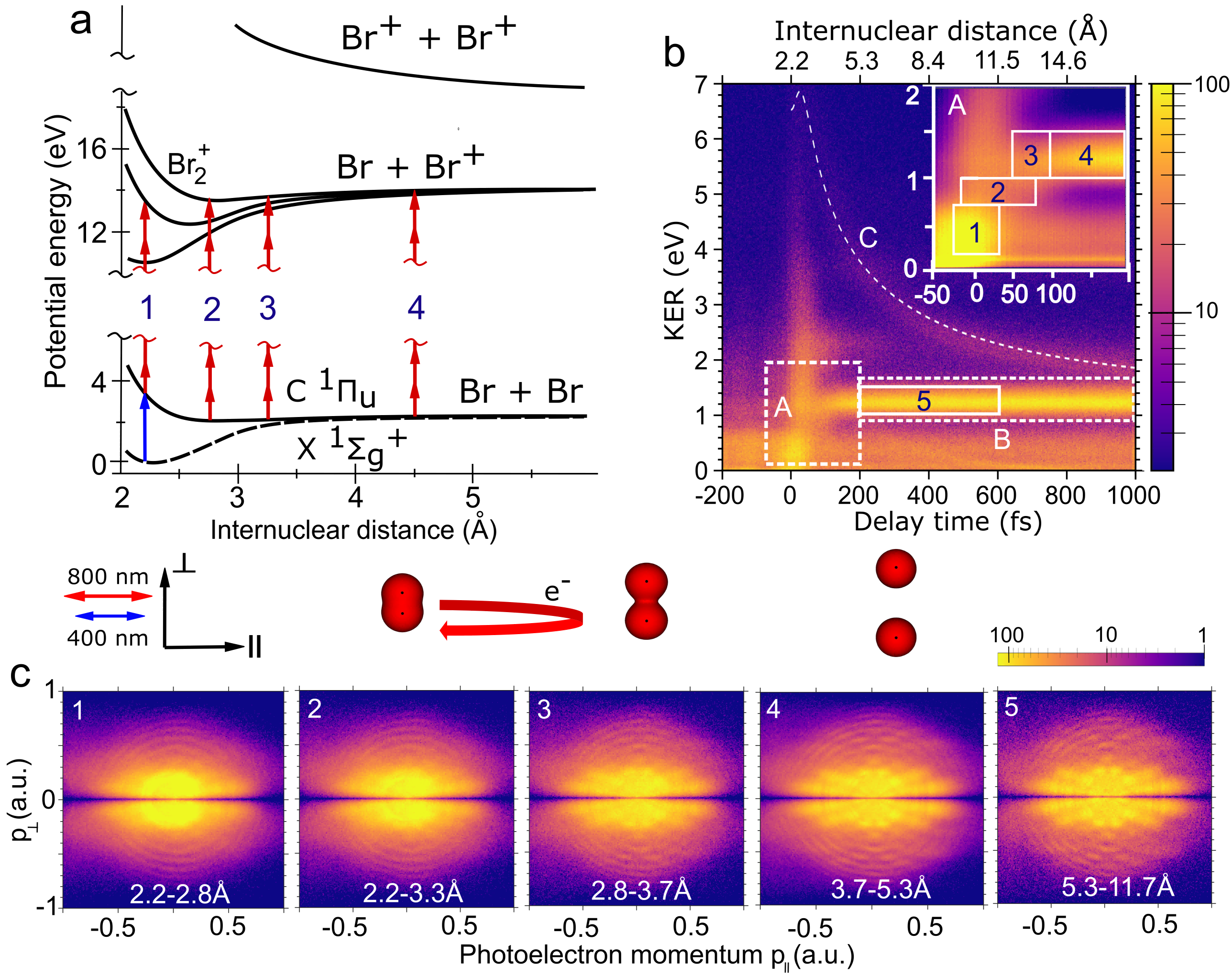}
    \caption{\textbf{Time-resolved tracking of nuclear and electronic dynamics during Br$_2$ dissociation.} \textbf{a,} Selected potential-energy curves of neutral and ionic Br$_2$ versus internuclear distance. A 400~nm pulse excites Br$_2$ from the ground state to the dissociative C state, initiating neutral dissociation into Br + Br. A time-delayed 800~nm pulse probes the evolving system through strong-field ionization at different internuclear separations (red arrows), accessing the single-ionization (Br + Br$^+$) and double-ionization (Br$^+$ + Br$^+$) channels. \textbf{b,} Kinetic-energy-release (KER) map reconstructed from the Br$^+$ recoil momentum as a function of pump-probe delay. Features A and B arise from single ionization at small and large internuclear separations, respectively, whereas feature C corresponds to Coulomb explosion from the doubly ionized Br$_2$. The inset defines KER-delay Zones 1--4; Zone 5 lies within feature B. The upper axis converts pump-probe delay into internuclear distance using a classical trajectory constrained by feature C (see SI \cite{supp}). The central schematic illustrates neutral dissociation and the strong-field ionization-recollision probe. The laser polarization defines the parallel and perpendicular directions; the molecular axis is preferentially aligned perpendicular to the polarization following photoexcitation. \textbf{c,} Photoelectron momentum distributions for the five KER-delay zones defined in \textbf{b}, displayed in the recoil-referenced $p_{\perp}$-$p_{\parallel}$ frame illustrated by the central schematic. Here, $p_{\parallel}$ is parallel to the laser polarization, while the direction of $p_{\perp}$ is referenced to the coincident Br$^+$ recoil. The corresponding internuclear-distance range is indicated in each panel.}
    \label{fig1}
\end{figure}

The time-resolved Br$^+$ kinetic-energy-release (KER) map in Fig.~\ref{fig1}b exhibits a pronounced low-KER feature, \textit{A}, emerging near zero delay and reaching an asymptote at around 200~fs and a KER of 1.2-1.3~eV. This feature originates from single ionization of the dissociating neutral molecule and traces the evolving electronic structure as the bond elongates \cite{Li2010PNAS}. As the atoms continue to separate, the KER stabilizes, giving rise to a delay-independent long tail (feature \textit{B}) that persists for more than 1~ps. At this stage, ionization predominantly probes isolated Br atoms because the interaction between Br and Br$^+$ becomes negligible at large internuclear separation. In principle, the tail would remain unchanged for several nanoseconds, until the dissociating molecules have left the laser focus. The inset of Fig.~\ref{fig1}b provides an expanded view of feature \textit{A}. For clarity, we divide this region of the KER–delay map into four representative zones 1--4, centered at the internuclear distances indicated in Fig.~\ref{fig1}a. The long-delay asymptotic region associated with feature B is treated separately as Zone 5. This segmentation is not unique, but is guided by changes in the KER distribution and the corresponding photoelectron momentum distributions, allowing us to distinguish different stages of dissociation without implying sharply defined boundaries. 

A weaker, high-KER feature, C, arises from double ionization and displays the characteristic Coulomb-explosion signature, with the KER decreasing approximately as $1 / R$ as the fragments separate. Assuming an approximately constant neutral-dissociation velocity over this delay range, this decrease maps onto the evolving internuclear distance $R$ relative to its equilibrium value, $R_e \approx 2.28~\mathrm{\AA}$. This mapping further assumes that the Br$_2^{2+}$ dication dissociates on an approximately Coulombic potential. The good agreement between the observed high-KER feature and the classical wave-packet simulation (white dashed curve in Fig.~\ref{fig1}b supports the validity of this approximation in the dissociation regime, where the interaction approaches the $1 / R$ limit, while deviations may still occur at shorter internuclear distances due to electronic-structure effects. Importantly, this illustrates the limitation of Coulomb-explosion imaging alone: it provides a measure of nuclear separation, but does not determine when the electronic structure has crossed from a molecular to an atomic regime.

To relate these nuclear dynamics to the accompanying electronic evolution, we analyze the photoelectrons detected in coincidence with Br$^+$ ions. Electrons correlated with each of the five KER-delay zones in Fig.~\ref{fig1}b are used to reconstruct molecular-frame photoelectron momentum distributions (PMDs) (see SI \cite{supp}). Fig.~\ref{fig1}c displays the PMDs in a signed cylindrical $p_{\perp},p_{\parallel}$ representation. Here, $p_{\parallel}$ is the photoelectron momentum component along the laser polarization and $|p_{\perp}|=\sqrt{p_x^2+p_y^2}$ is the radial momentum in the perpendicular plane. The azimuthal electron distribution is integrated separately over two half-planes defined relative to the transverse component of the coincident Br$^{+}$ recoil momentum; the sign of $p_{\perp}$ distinguishes electron emission into the parallel and antiparallel recoil-relative half-planes.

The PMDs evolve substantially over the course of the pump-probe delay. All PMDs exhibit above-threshold ionization (ATI) features in the form of radial rings \cite{PhysRevA.106.013106}, but the relative strength and angular structure of these features change as dissociation proceeds. At short delays, the low-momentum region is dominated by a pronounced inner ring, whereas at longer delays the ATI rings become increasingly modulated by angular structure and holographic interference features become more visible. 

The qualitative change in the PMDs already indicates a transition in the character of the ionization response as the system evolves from a molecule toward separated atoms. In atoms, the high symmetry of the ionic potential gives rise to comparatively clean angular partial-wave and interference structures in ATI spectra. In a dissociating molecule, by contrast, the outgoing electron is influenced by a two-center potential, a changing internuclear separation, and residual alignment averaging, which generally complicate and smear the angular structure. The emergence of more pronounced angular modulation and holographic interference at longer delays is therefore consistent with an increasingly atomic-like ionization response.

\begin{figure}[H] 
    \centering{\includegraphics[width=0.8\textwidth]{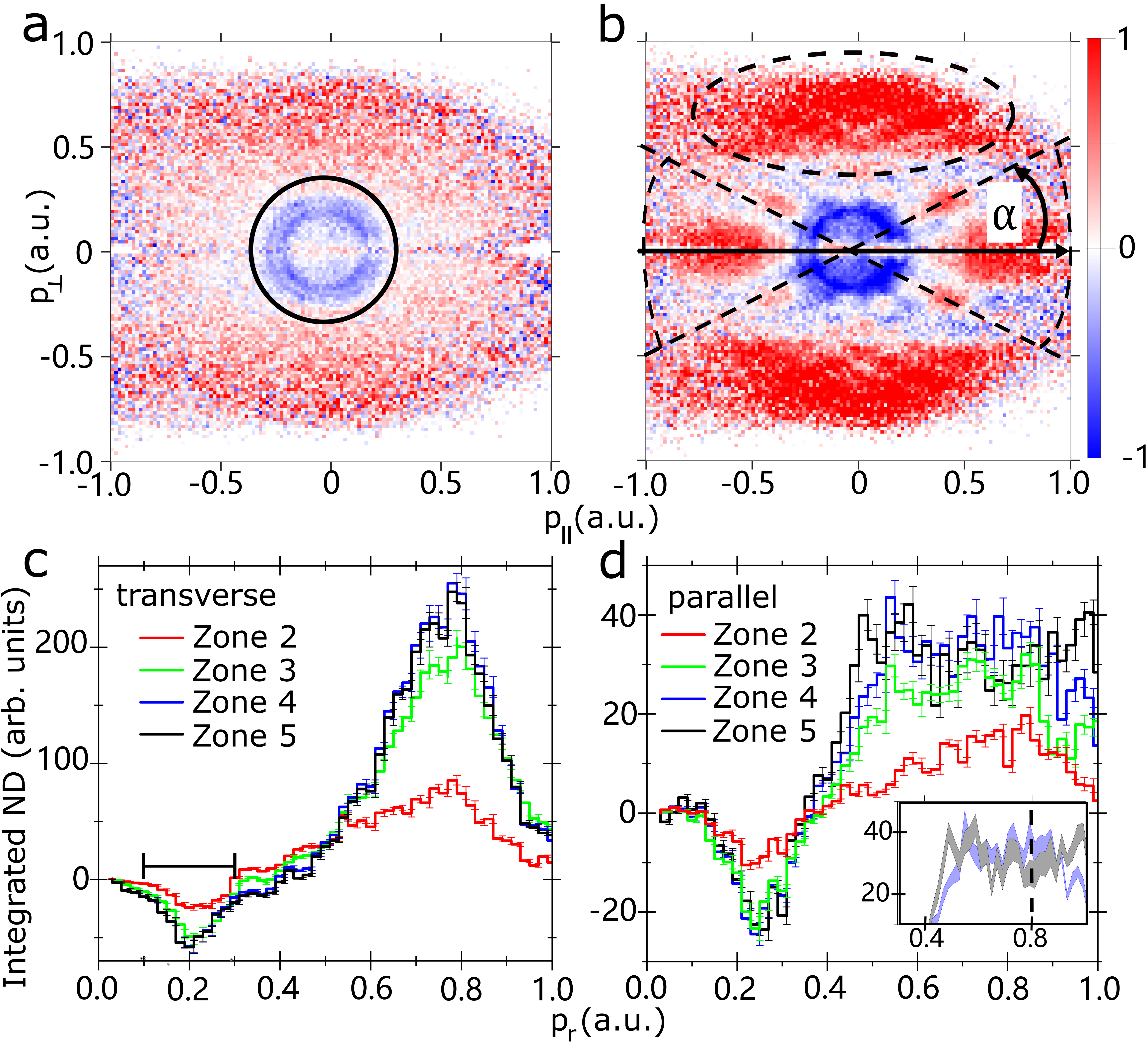}}
    \caption{\textbf{Evolution of photoelectron momentum during Br$_2$ dissociation.} \textbf{a,b}, ND-PMDs for Zone 2 and 4. The ND signal is defined relative to Zone 1 (ground-state molecule reference). The solid circle in \textbf{a} marks a low-momentum inner-ring feature, while the dashed ellipse and angular sector defined by dashed lines in \textbf{b} indicate regions of enhanced signal at higher momentum. \textbf{c,d}, Radial momentum distributions (RMDs) extracted from the ND-PMDs for Zones 2-5 along transverse (30$^{\circ}$ $<$ $\left|\alpha\right|$ $<$ 150$^{\circ}$) and parallel (($|\alpha| < 30^{\circ}$) or ($|\alpha| > 150^{\circ}$)) directions. The low-momentum inner-ring feature manifests as a valley around $p_r \approx 0.2~\mathrm{a.u.}$, marked by the horizontal black bar in \textbf{c}. In the parallel direction, a high momentum enhancement emerges over $p_r \approx 0.4$-$1.0~\mathrm{a.u.}$, with a clear separation between Zones 4 (blue) and Zone 5 (grey) at $p_r \approx 0.8~\mathrm{a.u.}$ (inset in \textbf{d}).}
    \label{fig2}
\end{figure}

The evolution of the PMDs is analyzed using normalized-difference (ND) PMDs, which highlight changes relative to a reference distribution. Using the Zone 1 PMD, $D_{\mathrm{ref}}(\mathbf{p})$, as the ground-state molecule reference, the ND-PMD for each subsequent zone is defined as: 
{\setlength{\abovedisplayskip}{4pt}
 \setlength{\belowdisplayskip}{4pt}
\begin{equation}
    ND(\textbf{p})=\frac{D_{sig}(\textbf{p})-D_{ref}(\textbf{p})}{D_{sig}(\textbf{p})+D_{ref}(\textbf{p})}
\end{equation}
}
 where $D_{\mathrm{sig}}(\mathbf{p})$ denotes the PMD for Zones 2--5. Representative ND-PMDs for Zones 2 and 4 are shown in Fig.~\ref{fig2}a, b, while those for all zones are provided in Figure S5 of the SI \cite{supp}.
 
Three distinct momentum-space features characterize the evolution of ND-PMDs: a low momentum negative (blue) inner ring, and two positive (red) high-momentum structures located predominantly in the transverse and parallel directions. To quantify their evolution, we extract radial momentum distributions (RMDs) by integrating the ND signal over angular sectors defined by the electron emission angle $\alpha$, measured relative to the laser polarization axis, as a function of the radial momentum $p_r=\sqrt{p_\parallel^2+p_\perp^2}$.

The parallel sector includes emission along either direction of the polarization axis ($|\alpha| < 30^{\circ}$ or $|\alpha| > 150^{\circ}$), while the transverse sector is defined by $30^{\circ} < |\alpha| < 150^{\circ}$ (Fig.~\ref{fig2}c,d).

The low-momentum inner ring appears as a valley centered at $p_r\approx0.2~\mathrm{a.u.}$ in the RMDs, reaching its minimum in Zone~3. The transverse high-momentum enhancement forms a pronounced peak, whereas the parallel enhancement appears as a broad plateau extending over $p_r\approx0.4$--$1.0~\mathrm{a.u.}$. The transverse peak approaches its limiting value in Zone~4, while the parallel plateau continues evolving between Zones~4 and~5 (inset of Fig.~\ref{fig2}d).

This angular-sector analysis is essential because the three dimensional momentum-space features evolve differently and would be obscured in a fully angle-integrated photoelectron distribution. We next use a semiclassical model to connect these distinct observables to orbital reshaping, charge localization, and the decay of the bond-breaking-state response.

\subsubsection*{Semiclassical model of the ionization dynamics}

To interpret the observed photoelectron features and identify their physical origin, we employ a two-dimensional semiclassical two-step (SCTS) model. Within this model, an electron tunnels through the instantaneous barrier formed by the combined molecular Coulomb and laser fields and is subsequently propagated classically in these fields, including rescattering \cite{PhysRevA.94.013415, PhysRevA.106.013106, Shilovski2021}. Each trajectory accumulates a quantum phase, and the coherent superposition of all trajectories gives rise to the measured photoelectron momentum distributions (see Methods and SI \cite{supp}). 

The SCTS model includes the laser field and the effective potential of the residual ion experienced by the departing electron. In the multicycle simulations, we use a laser field with constant amplitude over four optical cycles followed by a six-cycle turn-off. When ionization is restricted to the first optical cycle of this field, we refer to the results as single-cycle simulations \cite{supp}. The effective potential is parameterized by the effective ionic charge and ionic polarizability of the residual ion, which serve as a compact phenomenological representation to the evolving electronic structure during bond dissociation. The implementation of the SCTS model is described in the Methods and SI \cite{supp}. 

To establish the limiting electronic responses captured by the model, we first consider ionization of ground Br$_2$ and decoupled Br atoms. In these two limits, the electron density of the resulting ions is distributed over two or one atomic centers, respectively, and the response of the ion to the external laser field is characterized by its static polarizability, $\alpha_I(\mathrm{Br}_2^+,R_e)\approx40~\mathrm{a.u.^3}$ and $\alpha_I(\mathrm{Br}^+)\approx12~\mathrm{a.u.^3}$ (see SI \cite{supp}). This description breaks down as the Br–Br bond is stretched. As the internuclear separation increases, the bound electron density becomes increasingly diffuse and occupies a much larger spatial region, making the ionic electron cloud substantially easier to polarize by the external field, and thus giving rise to a pronounced enhancement of the ionic polarizability \cite{Hait2023}. Near the bond-breaking regime, but before the emergence of effectively decoupled atomic behavior, neither the molecular nor the atomic description alone adequately captures the evolving electronic response during dissociation. 

To describe this intermediate electronic-response regime within the SCTS model, we introduce the bond-breaking state (BBS), which is characterized by the coexistence of molecular and atomic characteristics arising from the partial loss of interatomic electronic coherence. Thus, we represent the electronic response of the BBS by two components: a localized atomic contribution associated with Br$^+$ [$\alpha_I(\mathrm{Br}^+)$], and a residual delocalized molecular contribution described by a stretched Br$_2^+$ configuration [$\alpha_I(\mathrm{Br}_2^+,2R_e)\approx80~\mathrm{a.u.^3}$] (estimated polarizability, see SI \cite{supp}). These two contributions are combined with equal weighting in the model (see Methods), providing a phenomenological description of the intermediate BBS. We then simulate the photoelectron momentum distributions for the molecular limit (ground-state Br$_2$), the intermediate BBS, and the atomic limit (decoupled Br atoms).

Having introduced the SCTS framework and the corresponding simulations, we now compare the calculated photoelectron momentum distributions with the experimental observations.

\begin{figure}[H] 
    \centering{\includegraphics[width=1.0\textwidth]{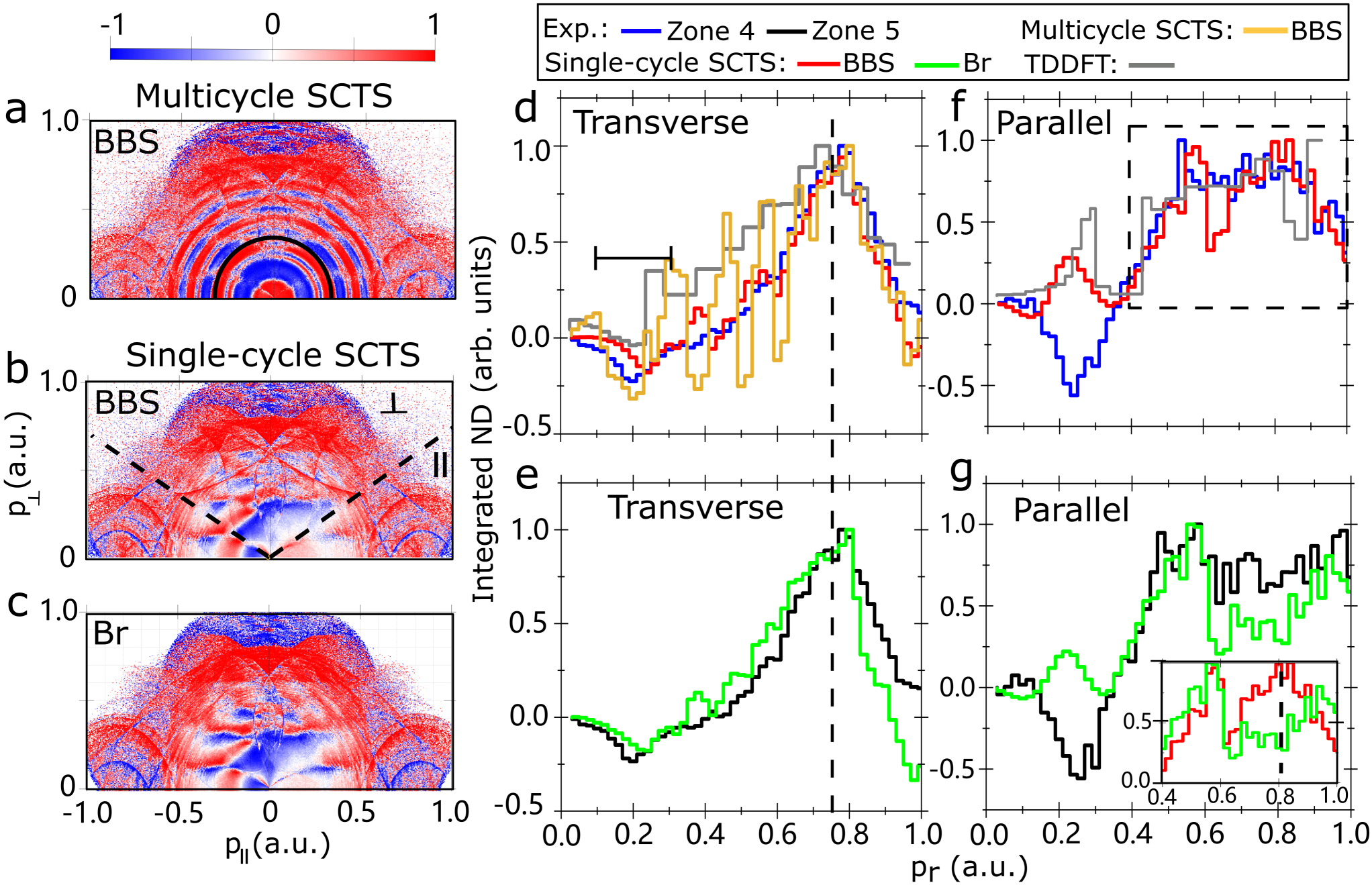}}
    \caption{\textbf{Photoelectron signatures of the bond-breaking state (BBS) and atomic limit.} \textbf{a--c,} Simulated ND-PMDs. \textbf{a,} Multicycle SCTS simulation of the BBS. Intercycle interference produces pronounced ATI rings; the black circle marks the low-momentum inner-ring feature. \textbf{b,c,} Single-cycle SCTS simulations of the BBS \textbf{b} and a decoupled Br atom \textbf{c}. Restricting ionization to one optical cycle suppresses intercycle ATI interference and emphasizes rescattering from the effective ionic potential. Dashed lines define the angular sectors used to extract the RMDs, as in Fig.~\ref{fig2}; each sector is normalized independently. \textbf{d,} Transverse RMDs from the single-cycle (red) and multicycle (yellow) BBS simulations, compared with experiment in Zone 4 (blue) and TDDFT calculations (grey). \textbf{e,} Transverse RMD from the single-cycle decoupled-Br simulation (green), compared with experiment in Zone 5 (black). The vertical dashed line marks the common transverse-peak position. \textbf{f,g,} Corresponding parallel RMDs for the BBS \textbf{f} and decoupled Br atom \textbf{g}. Zones 4 and 5 represent the intermediate BBS regime and long-delay atomic limit, respectively; the earlier evolution in Zones 1-3 is examined in Figs.~\ref{fig2} and \ref{fig4}. The dashed box and inset in \textbf{g} highlight the continued evolution of the parallel enhancement toward the atomic limit. The vertical dashed line in the inset marks the same radial momentum as in Fig.~\ref{fig2}d.}
    \label{fig3}
\end{figure}

\subsubsection*{Comparison of simulations and experiment}

Fig.~\ref{fig3}a shows the ND-PMD obtained from the multicycle BBS simulation, using the molecular PMD calculated at the equilibrium internuclear distance as the reference. The simulation qualitatively reproduces the main features of the experimental ND-PMDs: the low-momentum inner ring and the enhanced high-momentum signal in the transverse and parallel directions. The multicycle simulation also exhibits pronounced alternating ATI ring structures spaced by the photon energy, arising from intercycle interference between photoelectron trajectories \cite{PhysRevA.51.1495,PhysRevA.81.021403}.

To better distinguish scattering signatures from the dominant ATI rings, we restrict ionization in the model to a single optical cycle. This suppresses intercycle interference and emphasizes laser-driven rescattering, in which the liberated electron returns to the residual ion and scatters from its effective potential. The resulting high-momentum structures are therefore sensitive to the localized ionic charge and the surrounding polarizable electronic density, making them useful probes of the bond-breaking state. Fig.~\ref{fig3}b and c show the corresponding single-cycle ND-PMDs for the BBS and the decoupled Br atom, respectively.

The experimentally observed enhancement structures in the transverse and parallel sectors are reproduced by the single-cycle SCTS simulation of the BBS (Fig.~\ref{fig3}b). Quantitative comparisons between the simulations and experiment are provided by the RMDs extracted from the ND-PMDs in Fig.~\ref{fig3}d--g. Beyond the semiclassical description, we also perform quantum time-dependent density functional theory (TDDFT) simulations (see Methods and SI \cite{supp} for details).

In the transverse direction (Figs.~\ref{fig3}d,e), both the SCTS and TDDFT simulations reproduce the pronounced high-momentum peak observed experimentally in Zones~4 and~5, showing that this feature is robustly captured in both approaches. In the parallel direction (Fig.~\ref{fig3}f), however, only the BBS SCTS simulation reproduces the experimentally observed plateau-like enhancement, while the present TDDFT calculation shows substantially weaker agreement. 

The decoupled-Br simulation in Figs.~\ref{fig3}e,g instead represents the atomic response approached at longer delays. The inset in Fig.~\ref{fig3}g reveals a difference in the parallel high-momentum response between the BBS and decoupled-Br simulations, similar to the experimentally observed difference between Zones~4 and~5 in Fig.~\ref{fig2}d.

The weaker agreement of the TDDFT calculation in the parallel direction likely reflects the difficulty of standard approximate TDDFT approaches in describing the strongly correlated intermediate dissociation regime. In the BBS-based SCTS model, this regime is represented phenomenologically through the coexistence of localized atomic character and a residual delocalized electronic response during dissociation. The agreement between experiment and the BBS-based SCTS simulation therefore suggests that the parallel enhancement is closely connected to the residual delocalized electronic response associated with the bond-breaking state.

\subsubsection*{Interpretation of Results}

\begin{figure}[H] 
    \centering{\includegraphics[width=1.0\textwidth]{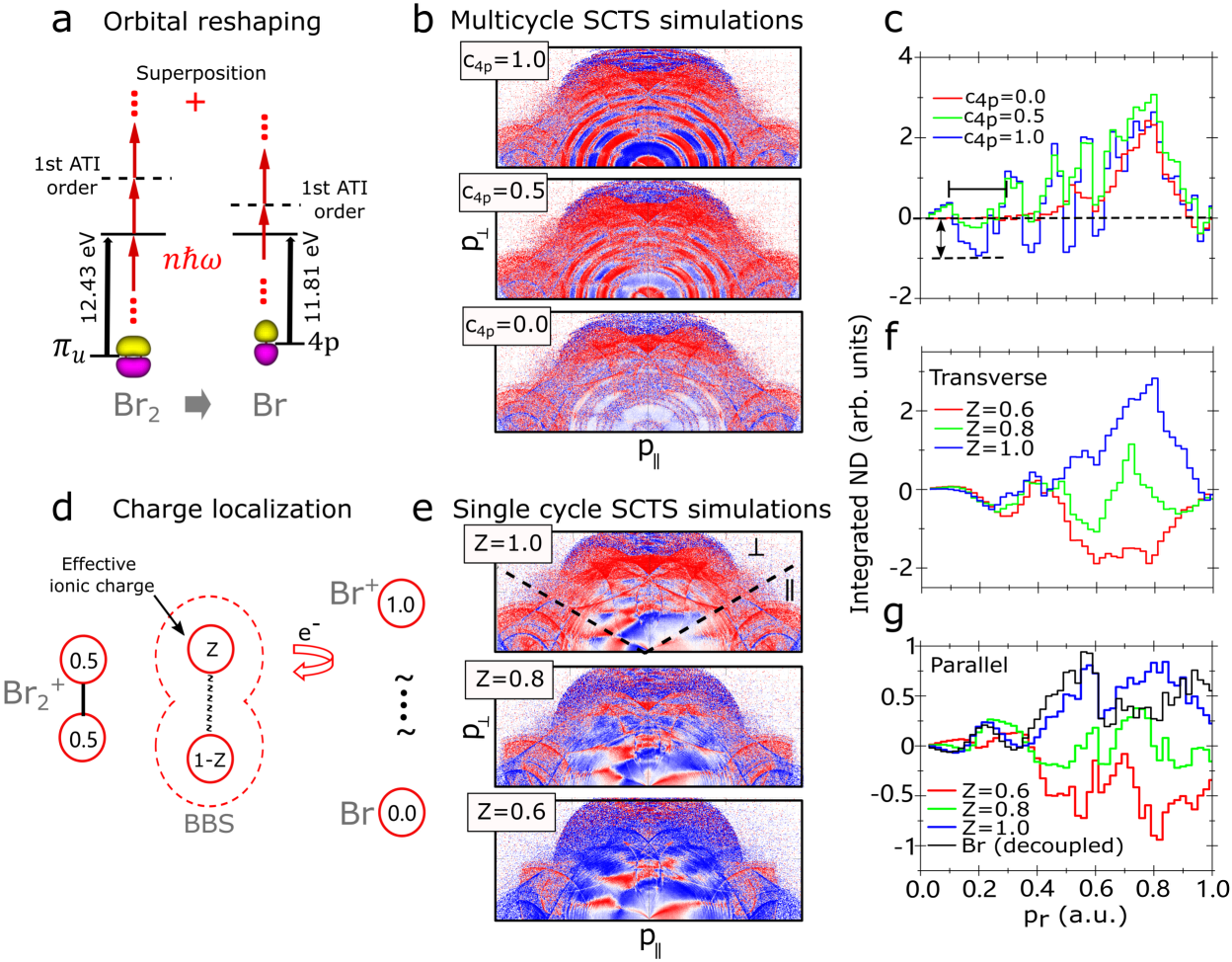}}
    \caption{\textbf{Simulated photoelectron signatures of the electronic evolution during bond dissociation.} \textbf{a}, Schematic illustration of the two-channel ATI process, in which the relative contributions of the molecular $\pi_u$ orbital and the atomic 4p orbital determine the ATI structure and probe the orbital reshaping during dissociation. \textbf{b,c,} Multicycle SCTS simulations illustrating the evolution of the ATI structure with varying relative atomic-orbital contribution $c_{4p}$. \textbf{b,} Normalized-difference (ND) maps. \textbf{c,} Corresponding radial momentum distributions (RMDs). \textbf{d,} Schematic illustration of the evolution of the effective ionic scattering potential experienced by the returning electron following strong-field ionization. As the internuclear distance increases, residual electronic coupling between the two Br centers gradually weakens, reducing charge sharing and continuously increasing the effective localized ionic charge from the molecular limit (0.5/0.5) toward the atomic limit (1.0/0.0). Within the bond-breaking-state (BBS), the returning electron experiences an intermediate scattering potential characterized by the effective localized ionic charge $Z$. \textbf{e--g,} Single-cycle SCTS simulations illustrating the evolution of the rescattering dynamics with varying effective localized ionic charge Z. \textbf{e,} ND-PMDs. \textbf{f,} Corresponding transverse RMDs. \textbf{g,} Parallel RMDs comparing the BBS and the decoupled Br atom, illustrating that the parallel enhancement continues evolving even after the localized ionic charge reaches the atomic limit, reflecting the gradual decay of the residual BBS response.}
    \label{fig4}
\end{figure}

Having shown that the BBS-based SCTS model reproduces the main experimental momentum structures, we next use controlled parameter variations to identify what each feature probes. In Fig.~\ref{fig4}, we vary the atomic Br 4p contribution to test the sensitivity of the ATI structure to the ionization channel and vary the effective ionic charge (Z) to test the sensitivity of the rescattering structures to charge localization. This separates the three signatures into complementary observables: the ATI-ring evolution tracks the effective ionization potential and dominant channel, the transverse high-momentum enhancement tracks localization of the scattering potential, and the parallel enhancement tracks the residual delocalized BBS response.

\textbf{Ionization-channel evolution and ATI structure.}
The evolution of the ATI structure reflects the changing ionization channel during dissociation. As illustrated in Fig.~\ref{fig4}a, strong-field ionization is dominated by the molecular $\pi_u$ orbital of Br$_2$ at small internuclear separations and evolves toward ionization from the atomic Br 4p orbital in the dissociation limit \cite{Li2010PNAS}. Because these two channels have different ionization potentials, they generate slightly shifted ATI combs in the photoelectron momentum distribution \cite{Andrey2012}. The gradual change in their relative contributions therefore modifies the ATI ring structure and provides a sensitive probe of the changing effective ionization potential and dominant ionizing-channel character.

To test this interpretation, we vary the normalized atomic-channel contribution, $c_{4p}$, in the multicycle SCTS simulations while keeping all other model parameters fixed. As $c_{4p}$ increases from the molecular to the atomic limit, the simulated ND-PMDs show a continuous evolution of the inner-ring structure (Fig.~\ref{fig4}b), accompanied by a shift and deepening of the low-momentum valley in the RMDs (Fig.~\ref{fig4}c). The same trend is observed experimentally across the KER-delay zones (Fig.~\ref{fig2}c), where the inner-valley structure deepens and stabilizes as dissociation proceeds. This agreement indicates that the ATI features trace the transition from molecular $\pi_u$-dominated ionization to atomic Br 4p-like ionization. The convergence of the ATI valley in Zone~3 shows that this ionization-channel evolution is well developed by approximately $75\pm25$~fs, consistent with previous photoelectron spectroscopy measurements \cite{Wernet2009PRL}.

\textbf{Charge localization and transverse enhancement.}
To identify the physical origin of the transverse enhancement, we examine how the rescattering dynamics depends on the spatial localization of the ionic scattering potential. In the molecular limit of Br$_2$, the residual ionic potential is distributed over two Br centers, corresponding in the model to two effective scattering centers of +0.5 (Fig.~\ref{fig4}d). In the atomic limit, on the contrary, the returning electron scatters from a localized single-center Br$^+$ potential with effective charge +1. A spatially distributed two-center potential produces weaker large-angle rescattering than a localized atomic potential because the Coulomb field experienced by the returning electron is spread over the internuclear region. As dissociation proceeds, the electronic structure evolves continuously from this delocalized molecular configuration toward a more atomic-like localized scattering potential.

Within the BBS framework, we describe this evolution phenomenologically through the effective localized ionic charge $Z$. This parameter should not be interpreted as a literal partial charge on one Br atom; after strong-field ionization, the residual ionic system carries a total charge of +1. Rather, $Z$ parameterizes the degree to which the scattering potential experienced by the returning electron has acquired a single-center, atomic-like character. In the intermediate bond-breaking regime, residual electronic coupling between the two Br centers keeps part of the Coulomb field delocalized over the molecule, reducing the effective localized scattering strength. As the fragments separate further, this delocalized contribution diminishes and the scattering potential approaches the atomic limit. We therefore vary $Z$ from representative intermediate values toward $Z=1$, while keeping all other model parameters fixed.

As $Z$ increases from 0.6 to 0.8 and 1.0, the transverse high-momentum enhancement in the simulated ND-PMDs becomes progressively stronger (Fig.~\ref{fig4}e), accompanied by a systematic increase of the corresponding transverse peak in the RMDs (Fig.~\ref{fig4}f). This model dependence shows that the transverse enhancement is sensitive to the localization of the effective scattering potential. The same qualitative trend is observed experimentally across the KER-delay zones (Fig.~\ref{fig2}c), where the transverse high-momentum enhancement increases from early delays toward Zones~4 and~5. The saturation of this transverse rescattering signature in Zone~4 is therefore consistent with the emergence of an atomic-like localized scattering potential by approximately $125\pm25$ fs.

\textbf{Bond-breaking-state decay and parallel enhancement.}
In contrast to the transverse scattering, the parallel enhancement evolves differently. In Fig.~\ref{fig3}f, the single-cycle BBS simulation reproduces the plateau-like parallel enhancement observed in Zone~4, while the decoupled-Br simulation (Fig.~\ref{fig3}g) shows a distinct high-momentum response corresponding to the atomic limit. Experimentally, the parallel RMD continues to evolve from Zone~4 to Zone~5, even as the transverse enhancement saturates (inset of Fig.~\ref{fig2}d). This indicates that the rescattering response has not yet reached the atomic limit when the localized scattering strength already appears atomic-like.

This difference does not imply direct imaging of separate density components. Instead, the transverse and parallel features probe the same evolving effective potential with different sensitivities. The transverse enhancement involves larger-angle scattering and is more sensitive to the localized ionic potential (parameterized by (Z)), while the parallel enhancement is dominated by near-axis trajectories and reflects long-range focusing and polarization effects. Their distinct evolution thus arises from different sensitivities to the changing potential during dissociation.

The continued evolution of the parallel enhancement from Zones~4 to~5 shows that, even after the localized potential becomes atomic-like, the effective potential for near-axis rescattering remains non-atomic. We attribute this to a residual polarizable, delocalized response of the bond-breaking state. As the nuclei separate, this response decays and the system approaches the behavior of two decoupled Br atoms, reflected in the transition of the parallel RMD from BBS-like to atomic-like.

Together, the transverse and parallel structures provide complementary insight into dissociation dynamics. The transverse enhancement tracks the emergence of a localized atomic-like potential, while the parallel enhancement remains sensitive to residual BBS response. Their differing evolution indicates that charge localization and the decay of the residual electronic response are not simultaneous. The continued change in the parallel enhancement suggests this residual response persists for ~250~fs after localization, implying a total dissociation timescale exceeding 300~fs, consistent with high-harmonic generation studies \cite{Worner2010}.

However, unlike the transverse high-momentum feature-which gradually approaches the peak structure expected for decoupled Br atoms as dissociation proceeds-the enhancement along the polarization axis continues to evolve beyond the BBS response (highlighted in the inset of Fig.~\ref{fig3}g).
\section*{Conclusion}
We have combined coincident photoion and photoelectron momentum measurements to follow Br$_2$ photodissociation in real time. The ion momenta determine the evolving internuclear separation, while distinct structures in the photoelectron momentum distributions reveal different stages of the accompanying electronic evolution. Supported by semiclassical modelling, the measurements identify an intermediate bond-breaking state in which localized atomic character coexists with a residual delocalized and polarizable electronic response.

Together, these observations establish a sequential evolution of electronic structure during bond dissociation. The ATI response shows that the dominant ionization channel evolves from molecular $\pi_u$-dominated to atomic 4p-like character by Zone~3, corresponding to approximately 75~fs. The transverse rescattering enhancement indicates that an atomic-like localized scattering potential becomes well developed by Zone~4, around 125-150~fs. The parallel enhancement, however, continues to evolve toward Zone~5, showing that the residual polarizable response of the bond-breaking state persists until approximately 400~fs, where the system approaches the response of two effectively decoupled Br atoms.

These results extend previous studies based on photoelectron spectroscopy \cite{Wernet2009PRL}, strong-field ion momentum spectroscopy \cite{Li2010PNAS,Rouzee2013JPB}, and high-harmonic generation spectroscopy \cite{Worner2010}, which separately identified signatures of orbital evolution, nuclear motion, or the emergence of atomic behavior. Here, the correlated momentum measurements resolve these signatures within a single experiment and establish their temporal ordering. The results show that different observables yield different characteristic dissociation times because they are sensitive to different stages of the coupled electron-nuclear dynamics. In this sense, the bond-breaking time is not a unique instant, but depends on which aspect of the evolving molecular-to-atomic transition is being probed.

\backmatter 
\section*{Methods}
\bmhead{Experiment} A two-color pump-probe setup based on a Mach-Zehnder interferometer seeded by a Ti:sapphire femtosecond laser system (800~nm, 25~fs, 10~kHz) was used to initiate and monitor Br$_2$ bond dissociation. In one arm of the interferometer, the 400~nm pump pulses were generated by frequency doubling the 800~nm pulses in a type-2 BBO (beta barium borate, $\beta$-BaB$_2$O$_4$)  crystal. The residual 800~nm light in this arm was removed from the 400~nm beam to better than $10^{-4}$ through two reflections on dichroic mirrors (each $10^{-2}$) and two reflections on custom-made chirped mirrors for 400~nm femtosecond pulses (compressed to 38~fs). In the second arm of the interferometer the 800~nm probe pulses were time-delayed using a motorized translation stage and their polarization was rotated to be parallel to the generated 400~nm light from the first interferometer arm. 
Both pulses were collinearly recombined on a dichroic mirror reflecting the 400~nm pulse and then steered onto a (normal-incidence) parabolic focusing mirror ($f=5$~cm) in an ultrahigh vacuum chamber ($10^{-10}$~mBar).
In the focus, the 400~nm ($\sim10^{10}$ W cm$^{-2}$) and 800~nm ($\sim5\times10^{13}$ W cm$^{-2}$) pulses intersected with a supersonic Br$_2$-He molecular jet (1\%~Br$_2$). Both pulses were linearly polarized in the same direction. Owing to the perpendicular transition character of the C-state excitation \cite{Cooper1998}, this configuration ensured that the neutral dissociation occurred perpendicular to the probe laser polarization.

\bmhead{Semi-classical two-step model} In our semiclassical model, strong-field ionization is described as a two-step process consisting of quantum tunneling followed by classical propagation of the ionized electron. The tunnel ionization rate is estimated by Ammosov-Delone-Krainov (ADK) rate $w_{ADK}$ \cite{Delone:91} and the tunneling exit point is approximated by the tunnel ionization in parabolic coordinates with induced dipole and Stark shift (TIPIS) approximation \cite{Pfeiffer2012NatPhys}. The subsequent electron propagation is governed by the system Hamiltonian:
\begin{equation*}
    H(t)=\frac{1}{2}\mathbf{p}^2+V(\mathbf{r})-\frac{\alpha_I\mathbf{F}\left(t\right)\cdot\mathbf{r}}{r^3}+\mathbf{r}\cdot\mathbf{F}(t),
\end{equation*}
where $V(\mathbf{r})$ is the Coulomb field of the parent ion, $\mathbf{F}(t)$ is the electric field of the laser pulse, and the static polarizability $\alpha_I$ depicts the interaction between the ionized electron and the multielectron parent ion. For the ground-state Br$_2$ ionization, $V(\mathbf{r})=-\frac{Z_{1}}{\left|\mathbf{r}-\mathbf{r_1}\right|}-\frac{Z_{2}}{\left|\mathbf{r}-\mathbf{r_2}\right|}$, where $\mathbf{r_1}=\left(0,-\frac{\mathbf{R}_e}{2}\right)$ and $\mathbf{r_2}=\left(0,\frac{\mathbf{R}_e}{2}\right)$ are defined by the equilibrium internuclear distance $\mathbf{R}_e$ of Br$_2$, and $Z_{1,2}=0.5$ represents the effective charge on each ion center of Br$_2^+$. For the single-centre BBS and decoupled-Br calculations, the Coulomb potential is represented as $V(\mathbf{r})=-\frac{Z}{\left|\mathbf{r}\right|}$.  The decoupled-Br atomic limit is calculated using (Z=1). For the controlled BBS calculations shown in Fig.~\ref{fig4}e,f, $Z$ is varied between 0.6, 0.8, and 1.0 while all other model parameters are held fixed. Here, $Z$ phenomenologically parameterizes the localized, single-centre scattering strength experienced by the returning electron and does not represent a literal partial charge on one Br atom. 

In the numerical implementation, a large ensemble of electron trajectories is generated using a Monte Carlo approach. For each trajectory, the ionization time \(t_0\) and the initial electron conditions \((\mathbf{v}_0,\mathbf{r}_0)\) are sampled, and the canonical equations derived from the Hamiltonian are integrated. To incorporate quantum interference between trajectories, the semiclassical phase \(\Phi(t_0,\mathbf{v}_0)\) is calculated as
\begin{equation*}
\begin{aligned}
    \Phi(t_0,\mathbf{v}_0)={}&-\mathbf{v}_0\cdot\mathbf{r}_0+I_p^{\mathrm{eff}}(t_0)t_0 \\
    &-\int_{t_0}^{\tau_f}\mathrm{d}t\left\{\frac{\mathbf{p}(t)^2}{2}+V[\mathbf{r}(t)]-\mathbf{r}(t)\cdot\boldsymbol{\nabla}
    \left[V[\mathbf{r}(t)]-\frac{\alpha_I\mathbf{F}(t)\cdot\mathbf{r}(t)}{|\mathbf{r}(t)|^3}\right]\right\}+\Phi_f^C(\tau_f),
\end{aligned}
\end{equation*}
where \((\mathbf{v}_0,\mathbf{r}_0)\) is the initial state of the tunnelling electron, \(\tau_f\) is the final propagation time, and \(\Phi_f^C(\tau_f)\) is the Coulomb phase \cite{PhysRevA.94.013415}. The phase contains the Stark-shifted effective ionization potential $I_p^{\mathrm{eff}}(t_0)=I_p+\frac{1}{2}\left(\alpha_N-\alpha_I\right)F^2(t_0)$, where \(I_p\) is the field-free ionization potential and \(\alpha_N\) and \(\alpha_I\) are the static polarizabilities of the neutral system and the residual ion, respectively.

At the end of the laser pulse, the asymptotic momentum of each electron is determined. The complex amplitudes of trajectories whose final momenta fall within the same two-dimensional momentum bin \(\mathbf{k}\) are summed coherently, resulting in the photoelectron momentum distribution
\begin{equation*}
    D(\mathbf{k}) = \left|S(\mathbf{k};\boldsymbol{\alpha},I_p)\right|^2,
\end{equation*}
where
\begin{equation*}
    S(\mathbf{k};\boldsymbol{\alpha},I_p)=\sum_{j\in\mathcal{B}(\mathbf{k})}\sqrt{w_{\mathrm{ADK}}\left[\mathbf{F}(t_0^j),I_p\right]} 
    \exp\left\{i\Phi^j\left[t_0^j,\mathbf{v}_0^j;\boldsymbol{\alpha},I_p\right]\right\},
    \qquad \boldsymbol{\alpha}=(\alpha_N,\alpha_I).
\end{equation*}
Here, \(\mathcal{B}(\mathbf{k})\) denotes the set of trajectories whose asymptotic momenta fall within the momentum bin \(\mathbf{k}\). The index \(j\) labels an individual trajectory with ionization time \(t_0^j\), initial velocity \(\mathbf{v}_0^j\), ADK weight \(w_{\mathrm{ADK}}^j\), and accumulated phase \(\Phi^j\).

Thus, \(t_0\) and \(\mathbf{v}_0\) are trajectory variables that are sampled and coherently summed over, whereas \(I_p\) and \(\boldsymbol{\alpha}\) characterize the electronic configuration used in the simulation. For the fixed laser pulse considered here, comparisons between different electronic configurations are performed by changing their corresponding ionization potentials and polarizabilities while sampling the ionization times over the same laser field \cite{Shilovski2016PRA,PhysRevA.106.013106}.

\bmhead{Bond-breaking state (BBS)}

In the semiclassical model, the bond-breaking state (BBS) introduced in the main text is represented phenomenologically as the coherent superposition of an atomic-like and a molecular-like electronic response. The atomic component is described using the polarizability of Br$^+$, whereas the molecular component is represented by a stretched Br$_2^+$ configuration at an internuclear separation of $2R_e$. Representative polarizabilities for the two components are adopted from literature values (see Supplementary Information for details).

The BBS represents an intermediate dissociation regime in which atomic-like and molecular-like electronic responses coexist. The total photoelectron amplitude is constructed as the coherent sum of two trajectory ensembles,
\begin{equation}
S_{\mathrm{BBS}}\left(I_p,N\right)
=
S_1\left(\pmb{\alpha}(\mathrm{Br}),I_p,N_1\right)
+
S_2\left(\pmb{\alpha}(\mathrm{Br}_2,2R_e),I_p,N_2\right),
\end{equation}
where
\[
S_1=\sum_{j=1}^{N_1}I^j\left(\pmb{\alpha}(\mathrm{Br}),I_p^j\right),\qquad
S_2=\sum_{j=1}^{N_2}I^j\left(\pmb{\alpha}(\mathrm{Br}_2,2R_e),I_p^j\right).
\]

Equal weighting is imposed by setting $N_1=N_2=0.5N$, with a total of $N=5\times10^8$ trajectories.

The same numerical procedure is applied to ground-state Br$_2$, isolated Br, and the BBS. Additional calculations with varied physical parameters are described in the Supplementary Information \cite{supp}.

\bmhead{Simulation of Ionization-channel evolution}To investigate how Ionization-channel evolution influences the photoelectron momentum distributions, we extend the coherent summation of the BBS model by introducing independent trajectory ensembles associated with ionization from the molecular $\pi_u$ orbital and the atomic 4p orbital. The total semiclassical amplitude is written as:
\begin{equation*}
    S_{BBS}^{\prime}=S_{\mathrm{BBS}}\left(I_p\left(\pi_u\right), N_{\pi_u}\right)+S_{\mathrm{BBS}}\left(I_p\left(4p\right), N_{4p}\right)
\end{equation*}
where $N_{\pi_u}$ and $N_{4p}$ denote the numbers of trajectories assigned to the two orbital channels, respectively. Since each trajectory is independently weighted by the ADK ionization rate according to its corresponding ionization potential, varying the relative trajectory numbers changes only the relative orbital contributions while leaving the underlying tunneling dynamics unchanged. The normalized atomic-orbital contribution is therefore defined as:
\begin{equation*}
    c_{4p}=\frac{N_{4p}}{N_{4p}+N_{\pi_u}},
\end{equation*}
which varies continuously from 0 (purely molecular $\pi_u$) to 1 (purely atomic 4p). The full simulated photoelectron momentum distributions with different $c_{4p}$ are given in the SI \cite{supp}.

\bmhead{Time-dependent density functional theory (TDDFT)}The TDDFT calculations are performed using Octopus software package~\cite{marques2003, tancogne-dejean2020}, which solves the time-dependent Kohn-Sham equations \cite{gross2012} on a real-spaced grid. The grid is 200 x 200 x 200 atomic units, with a spacing of 0.5 a.u., and we use a $\sin^{2}$ complex absorbing potential that extends 30 a.u.\ from each edge of the simulation box. We use a local-density-approximation (LDA) exchange-correlation functional~\cite{perdew1981} and an average-density self-interaction correction (ADSIC)~\cite{tsuneda2014}, and the bromine atoms are each described by a Hartwigsen–Goedecker–Hutter-type pseudopotential~\cite{hartwigsen1998}.

For a given internuclear distance $d$, we begin by computing the $C$ state of the Br$_{2}$ molecule. This is done by removing one electron from the $\pi_{g}^{*}$ Kohn-Sham orbital and placing it into the $\sigma_{u}^{*}$ orbital, and performing a ground-state DFT calculation in this state. We then apply a 5-cycle, $800\, \text{nm}$, $10\, \text{TW}/\text{cm}^{2}$ laser pulse to ionize the system. The outgoing photoelectron density is recorded using the time-dependent surface-flux (t-SURFF) method~\cite{scrinzi2012}, implemented in Octopus. We use a cubical surface with a side length of 120 atomic units for this t-SURFF calculation.

The corresponding original photoelectron momentum distributions and normalized difference (ND) analysis are provided in the SI \cite{supp}.






\bibliography{Bibliography}

\end{document}